\documentclass[journal]{IEEEtran}
\usepackage{cite}
\usepackage{amsfonts,dsfont,amsthm,amsmath,amssymb,amscd,epsfig,bm,subfigure,graphicx,mathtools}
\usepackage{textcomp}
\usepackage[utf8]{inputenc}
\usepackage[T1]{fontenc}
\usepackage{tabularx} 
\usepackage{multirow}
\usepackage{wasysym}
\usepackage{algorithm}
\usepackage{algorithmic}
\usepackage{suffix}
\usepackage{url}
\usepackage{multicol}
\usepackage{verbatim}

\usepackage[nolist]{acronym}
\usepackage{color,soul}
\usepackage[switch]{lineno}

\newcommand{\blue}[1]{\textcolor{black}{#1}}

\usepackage{xcolor}

\IEEEoverridecommandlockouts

\begin{document}
\title{Containing Future Epidemics with \\Trustworthy Federated Systems for Ubiquitous Warning and Response}

 \author{
 Dick~Carrillo,~
 Lam~Duc~Nguyen,~
 Pedro H. J. Nardelli,~
 Evangelos Pournaras,~
 Plinio Morita,~
 Demóstenes Z. Rodríguez,~
 Merim Dzaferagic,
 Harun Siljak,~
 Alexander Jung,~
 Laurent Hébert-Dufresne,~
 Irene Macaluso,~
 Mehar~Ullah,~
 Gustavo Fraidenraich,~
 Petar Popovski~
 \thanks{DC, PHJN and MU are with LUT University, Finland.
 LDN and PP is with Aalborg university, Denmark.
 EP is with University of Leeds, Leeds, UK,
 PM is with university of Waterloo, Canada,
 DZR is with federal university of Lavras, Brazil,
 MD, HS and IM are with Trinity College Dublin, Ireland,
 AJ is with Aalto University, Finland,
 LHD is with University of Vermont, USA,
 GF is with State University of Campinas, Brazil,
 This paper is partly supported by Academy of Finland via EE-IoT project n.319009.
 Part of this manuscript has been released as a pre-print at LUTPub \cite{preprintLUT}.
Contact: dick.carrillo.melgarejo@lut.fi.}
 }

\maketitle
\begin{abstract}
In this paper, we propose a  global digital platform to avoid and combat epidemics by providing relevant real-time information to support selective lockdowns.
It leverages the pervasiveness of wireless connectivity while being trustworthy and secure.
The proposed system is conceptualized to be decentralized yet federated, based on ubiquitous public systems and active citizen participation. Its foundations lie on the principle of informational self-determination.
We argue that only in this way it can become a trustworthy and legitimate public good infrastructure for citizens by balancing the asymmetry of the different hierarchical levels within the federated organization while providing highly effective detection and guiding mitigation measures towards graceful lockdown of the society.

To exemplify the proposed system, we choose \blue{a} remote patient monitoring as use case. \blue{This use case is evaluated considering different number of endorsed peers on a solution that is based on the integration of distributed ledger technologies and NB-IoT (narrowband IoT)}.
An experimental setup is used to evaluate the performance of this integration, in which the end-to-end latency is slightly increased when a new endorsed element is added. However, the system reliability, privacy, and interoperability are guaranteed.
In this sense, we expect active participation of empowered citizens to supplement the more usual top-down management of epidemics.
\end{abstract}


\begin{acronym}
  \acro{1G}{first generation of mobile network}
  \acro{1PPS}{1 pulse per second}
  \acro{2G}{second generation of mobile network}
  \acro{3G}{third generation of mobile network}
  \acro{4G}{fourth generation of mobile network}
  \acro{5G}{fifth generation}
  \acro{ARQ}{automatic repeat request}
  \acro{ASIP}{application specific integrated processors}
  \acro{AWGN}{additive white Gaussian noise}
   \acro{BER}{bit error rate}
  \acro{BCH}{Bose-Chaudhuri-Hocquenghem}
  \acro{BRICS}{Brazil-Russia-India-China-South Africa}
  \acro{BS}{base station}
  \acro{CDF}{cumulative density function}
  \acro{CoMP} {cooperative multi-point}
  \acro{CP}{cyclic prefix}
  \acro{CR}{cognitive radio}
  \acro{CS}{cyclic suffix}
  \acro{CSI}{channel state information}
  \acro{CSMA}{carrier sense multiple access}
  \acro{DFT}{discrete Fourier transform}
  \acro{DFT-s-OFDM}{DFT spread OFDM}
  \acro{DSA}{dynamic spectrum access}
  \acro{DVB}{digital video broadcast}
  \acro{DZT}{discrete Zak transform}
  \acro{eMBB}{enhanced mobile broadband}
  \acro{EPC}{evolved packet core}
  \acro{FBMC}{filterbank multicarrier}
  \acro{FDE}{frequency-domain equalization}
  \acro{FDMA}{frequency division multiple access}
  \acro{FD-OQAM-GFDM}{frequency-domain OQAM-GFDM}
  \acro{FEC}{forward error control}
  \acro{F-OFDM}{Filtered Orthogonal Frequency Division Multiplexing}
  \acro{FPGA}{field programmable gate array}
  \acro{FTN}{faster than Nyquist}
  \acro{FT}{Fourier transform}
  \acro{FSC}{frequency-selective channel}
  \acro{GFDM}{generalized frequency division multiplexing}
  \acro{GPS}{global positioning system}
  \acro{GS-GFDM}{guard-symbol GFDM}
  \acro{IARA}{Internet Access for Remote Areas}
  \acro{ICI}{intercarrier interference}
  \acro{IDFT}{Inverse Discrete Fourier Transform}
  \acro{IFI}{inter-frame interference}
  \acro{i.i.d.}{independent and identically distributed}
  \acro{IMS}{IP multimedia subsystem}
  \acro{IoT}{Internet of Things}
  \acro{IP}{Internet Protocol}
  \acro{ISI}{intersymbol interference}
  \acro{IUI}{inter-user interference}
  \acro{LDPC}{low-density parity check}
  \acro{LLR}{log-likelihood ratio}
  \acro{LMMSE}{linear minimum mean square error}
  \acro{LTE}{Long-Term Evolution}
  \acro{LTE-A}{Long-Term Evolution - Advanced}
  \acro{M2M}{Machine-to-Machine}
  \acro{MA}{multiple access}
  \acro{MAR}{mobile autonomous reporting}
  \acro{MF}{Matched filter}
  \acro{MIMO}{multiple-input multiple-output}
  \acro{MMSE}{minimum mean squared error}
  \acro{MRC}{maximum ratio combiner}
  \acro{MSE}{mean-squared error}
  \acro{MTC}{Machine-Type Communication}
  \acro{NEF}{noise enhancement factor}
  \acro{NFV}{network functions virtualization}
  \acro{OFDM}{orthogonal frequency division multiplexing}
  \acro{OOB}{out-of-band}
  \acro{OOBE}{out-of-band emission}
  \acro{OQAM}{offset quadrature amplitude modulation}
  \acro{PAPR}{peak-to-average power ratio}
  \acro{PDF}{probability density function}
  \acro{PHY}{physical layer}
  \acro{QAM}{quadrature amplitude modulation}
  \acro{PSD}{power spectrum density}
  \acro{QoE}{quality of experience}
  \acro{QoS}{quality of service}
  \acro{RC}{raised cosine}
  \acro{RRC}{root raised cosine}
  \acro{RTT} {round trip time}  
  \acro{SC}{single carrier}
  \acro{SC-FDE}{Single Carrier Frequency Domain Equalization}
  \acro{SC-FDMA}{Single Carrier Frequency Domain Multiple Access}
  \acro{SDN}{software-defined network}
  \acro{SDR}{software-defined radio}
  \acro{SDW}{software-defined waveform}
  \acro{SEP}{symbol error probability}
  \acro{SER}{symbol error rate}
  \acro{SIC}{successive interference cancellation}
  \acro{SINR}{signal-to-interference-and-noise ratio }
  \acro{SMS}{Short Message Service}
  \acro{SNR}{signal-to-noise ratio}
  \acro{STC}{space time code}
  \acro{STFT}{short-time Fourier transform}
  \acro{TD-OQAM-GFDM}{time-domain OQAM-GFDM}
  \acro{TTI}{time transmission interval}
  \acro{TR-STC}{Time-Reverse Space Time Coding}
  \acro{TR-STC-GFDMA}{TR-STC Generalized Frequency Division Multiple Access}
  \acro{TVC}{ime-variant channel}
  \acro{UFMC}{universal filtered multi-carrier}
  \acro{UF-OFDM}{Universal Filtered Orthogonal Frequency Multiplexing}
  \acro{UHF}{ultra high frequency}
  \acro{URLL}{Ultra Reliable Low Latency}
  \acro{V2V}{vehicle-to-vehicle}
  \acro{V-OFDM}{Vector OFDM}
  \acro{ZF}{zero-forcing}
  \acro{ZMCSC}{zero-mean circular symmetric complex Gaussian}
  \acro{W-GFDM}{windowed GFDM}
  \acro{WHT}{Walsh-Hadamard Transform}
  \acro{WLAN}{wireless Local Area Network}
  \acro{WLE}{widely linear equalizer}
  \acro{WLP}{wide linear processing}
  \acro{WRAN}{Wireless Regional Area Network}
  \acro{WSN}{wireless sensor networks}
  \acro{ROI}{return on investment}
  \acro{NR}{new radio}
  \acro{SAE}{system architecture evolution}
  \acro{E-UTRAN}{evolved UTRAN}
  \acro{3GPP}{3rd Generation Partnership Project }
  \acro{MME}{mobility management entity}
  \acro{S-GW}{serving gateway}
  \acro{P-GW}{packet-data network gateway}
  \acro{eNodeB}{evolved NodeB}
  \acro{UE}{user equipment}
  \acro{DL}{downlink}
  \acro{UL}{uplink}
  \acro{LSM}{link-to-system mapping}
  \acro{PDSCH}{physical downlink shared channel}
  \acro{TB}{transport block}
  \acro{MCS}{modulation code scheme}
  \acro{ECR}{effective code rate}
  \acro{BLER}{block error rate}
  \acro{CCI}{co-channel interference}
  \acro{OFDMA}{orthogonal frequency-division multiple access}
  \acro{LOS}{line-of-sight}
  \acro{VHF}{very high frequency}
  \acro{pdf}{probability density function}
  \acro{ns-3}{Network simulator 3}
  \acro{Mbps}{mega bits per second}
  \acro{EH}{energy harvesting}
  \acro{SWIPT}{simultaneous wireless information and power transfer}
  \acro{AF}{amplify-and-forward}
  \acro{DF}{decode-and-forward}
  \acro{WIT}{wireless information transfer}
  \acro{WPT}{wireless power transfer}
  \acro{FSFC}{frequency selective fading channel}
  \acro{DC}{direct current}
  \acro{FFT}{fast Fourier transform}
  \acro{RF}{radio frequency}
  \acro{SISO}{single-input single-output}
  \acro{RRC}{root raised cosine}
  \acro{TSR}{time-switching relaying}
  \acro{IFFT}{inverse fast Fourier transform}
  \acro{LIS}{large intelligent surfaces}
  \acro{URLLC}{ultra-reliable low-latency communication}
  \acro{ZMCSCG}{zero mean circularly symmetric complex Gaussian}
  \acro{PPSINR}{post-processing SINR}
  \acro{mMTC}{massive machine-type communication}
  \acro{NR}{New radio}
  \acro{RIS}{reconfigurable intelligent surface} 
  \acro{RAN}{radio access network}
  \acro{i.i.d.}{independent and identically distributed}
  \acro{NOMA}{non-orthogonal multiple access}
  \acro{SDN}{software defined networks}
  \acro{EMC}{edge-mobile computing}
  \acro{D2D}{device-to-device}
  \acro{LPWAN}{low-power wide-area network}
  \acro{WAN}{wide-area network}
  \acro{PAN}{personal-area network}
  \acro{BLE}{Bluetooth low energy}
  \acro{NB-IoT}{narrowband Internet of things }
 
  \acro{COVID-19}{coronavirus disease}
  \acro{WHO}{world health organization}
  \acro{CAPEX}{capital expenses}
  \acro{OPEX}{operational expenses}
  \acro{GEEWS}{global epidemiological early warning and response systems}
  \acro{GDP}{gross domestic product}
  \acro{DLT}{distributed ledger technology}
  \acro{NB-IoT}{narrowband IoT}
  \acro{KPI}{key performance indicator}
\end{acronym}

\section{Introduction} 

The COVID-19 pandemic has clearly shown that, in many senses, the world as a whole was not prepared for dealing with {a disease} of such magnitude. 
Although a final assessment is infeasible at this point (October, 2020), in which a second wave is creeping back in Europe and is poised to rage across the continent by fall 2020~\cite{Cacciapaglia2020_secondwave}.
The current statistics available about COVID-19 indicate that the most 
successful policies for monitoring and controlling the virus propagation are employing various digital technologies and connectivity~\cite{ting2020digital,ienca2020responsible}. 
These technologies can be used in two principal ways: (1) to provide status and predictions of the  epidemiological spread and facilitate actions, such as administration of diagnostic tests or preparation of medical equipment; (2) to implement active policies that facilitate societal processes and safe citizen movement, such as shop schedules, organized delivery of goods, and similar. 
The success of such informed policies is also strongly related to the 
\textit{timing of their implementation} for the respective place, as virus contagion 
is a spatiotemporal phenomenon that usually leads to exponential growth in the number of infected 
individuals \cite{adam2020special}.

Unfortunately, even the most successful timely interventions are in some 
way overreaching, shutting down abruptly most of the economic activities and creating social distancing between citizens.
While acknowledging that this is our current societal need to contain the propagation of this pandemic, our main hypothesis is that, in the near future, digital technologies and wireless connectivity can enable a \emph{graceful lockdown} by having the following main roles.
\begin{itemize}
    \item Facilitate selective lockdowns by allowing flexible transfer of the work, learn and play activities online to the desired extent.
    \item Assist in maintaining physical distancing and tracking the spreading of the disease, while offering digital tools to control the level of lockdown/reopening.
    \item Provide real-time information for targeted and efficient testing.
\end{itemize}
Note that these are also important today (October, 2020), when a second wave is forcing several countries to rethink a step-by-step reopening based on the aforementioned points. 

This will be achieved based on an integrated solution for the ubiquitous and privacy-preserving data acquisition, explainable predictive methods for contagion risk assessment, and digitized policies for selective lockdown and scheduling of societal activities.
Data will be locally acquired via online social networks, personal smart phone devices and Internet-of-Things (IoT) sensors in general.
It is foreseen that an effective global solution shall be decentralized, yet hierarchical (system-of-systems), to provide fine-tuned coordination among the federated entities to orchestrate actions
and then avoid {both the human rights cost of overly stringent solutions \cite{kupferschmidt2020can} and the human life cost of ``hands-off'' approaches.}
Beyond more traditional centralized policy-making, such a decentralized platform, which will be built using trustworthy \ac{DLT}, should also provide citizens with tools for more direct participation in mitigation measures based on tailored individual incentives.

In this sense, we expect to bring managerial decisions closer to the citizens.
With such a federated governance, we attempt to convert the usually passive data acquisition process to an active participatory one.
Hence, our aim is to move away from a purely top-down management approach.
In addition, the proposed platform aims at achieving the minimal use of personal data, processed with the maximum possible privacy-preservation to eliminate global health risks, while protecting the economy~\cite{Asikis2018}.
The COVID-19 pandemic has demonstrated that nations and governments need to be better prepared {and coordinated} to detect and react to global threats focusing on \textit{early detection, early response}, moving beyond traditionally passive disease monitoring based on voluntary reporting systems.
Hence, a rapid, participatory responsive detection system must be in place to support public health officials. A synergistic relationship with public health officials, policymakers, and citizens' active participation will be critical to align the mandate for public health with the protection of privacy, freedom and democracy~\cite{Helbing2019,Amat2020}.
In this context, one main factor is to design a special set of incentives that would allow the citizens to provide secured anonymized access to their data while actively participating in the crowd platform to support early disease detection, {a public information system}, and possible mitigation measures.
Building up such an incentive structure that maps those trade-offs is a key aspect (beyond purely technical ones) for the success of the proposed federated architecture. Furthermore, this system should be constructed and evaluated rigorously following ethical guidelines as in  \cite{ienca2020responsible,Hagendorff2020}.

Our contribution in this paper is threefold: 
\begin{enumerate}%
    \item a comprehensive analysis of epidemiological models, data collection, and wireless connectivity is done based on key relevant scientific references on pandemics.
    \item a federated global epidemiological warning system is proposed based on \acp{DLT}.
    \item an experimental setup of the integration between DLT and NB-IoT is used to evaluate the wireless network performance on the IoT infrastructure supporting a remote patient monitoring use case. 
\end{enumerate}%

The rest of the paper is organized as follows: 
Section II presents a discussion of epidemiological models and their limitations. 
Section III describes the relevance of wireless connectivity on pandemic scenarios.
In Section IV, the federated global epidemiological warning system is proposed, here is also detailed the experimental setup of the integration of \ac{DLT} and \ac{NB-IoT}.
Section V is reserved for some discussion and future perspectives.

\section{Epidemiological Models and Data Collection}

Biomedical data alone do not contain the information required for preventing and mitigating pandemics. In a highly interconnected globalized society, social interactions, environmental data, spatiotemporal events, and collective nonlinear phenomena observed on complex infrastructures such as transport systems, require the mining of heterogeneous pandemic big data that are a result of a complex system of systems processes.
In fact, modern disease monitoring systems, as the Global Public Health Intelligence Network (GPHIN) \cite{mykhalovskiy2006global} and the Global Outbreak Alert and Response Network (GOARN) \cite{davies2016gphin},  and event-based  approaches that use a combination of web crawlers, artificial intelligence, and public health expertise in the detection of indicators \cite{carter2018revitalizing,bernard2018intelligence} provide good---although limited---examples of successful platforms to identify latent indicators of an outbreak.

To go beyond those existing solutions, data scale, contextualization, and granularity are key requirements for data quality, which is usually orthogonal to privacy preservation, timely processing and analysis as well as storage and processing cost. 
\blue{In the following section}, more details will be provided about existing epidemiological models and epidemic-related data collection.

\subsection{Models and their limitations}

Models for emerging epidemics come in various forms, all relying on differently coarse-grained individuals’ data. State-of-the-art population-level models for COVID-19 rely on metapopulation models with an underlying Susceptible-Exposed-Infectious-Recovered (SEIR) disease dynamics \cite{chinazzi2020effect}. 
The metapopulations represent regions surrounding travel hubs and are interconnected through mobility data, for example, data from airlines, public transportation systems, and traffic control systems. 
The disease dynamics itself accounts for the fact that exposed individuals can be undergoing an incubation period, or even presymptomatic and unaware that they could infect others, as suggested by early case data\cite{wei2020presymptomatic}.
Note that some models opt for a simpler formulation~\cite{Zhang1481}, the Susceptible-Infectious-Recovered (SIR) dynamics which ignore the exposed state and therefore eliminate a potentially important delay in disease progression.

\blue{Unfortunately, while SEIR and SIR models can be } effective at providing large-scale forecasts and predicting importations at aggregated levels, \blue{ they are less informative at the individual level where we expect significant heterogeneity across individuals; both in behavior and disease manifestation \cite{10.1371/journal.pbio.3000897}.
One key assumption of these models is that all individuals of a given state follow the same dynamics (where state can be, for example, a combination of age and epidemiological status). However, the importance of superspreading events have highlighted that pandemics are often not driven by the average individual but by rare, stochastic, events where a synergy between individual behavior and disease manifestation can significantly spread the disease \cite{wang_inference}. }
\blue{Models informed by data at the individual level attempt to account for stochastic events but are less used for forecasting purposes as the data required are seldom available and hard to connect across multiple scales or populations \cite{dufresne_infections}. However, individual contact data can inform network models which are powerful for other purposes: Such as informing or measuring the impact of interventions and policies, such as school closures and lockdowns \cite{Zhang1481}. Likewise, these network models can be instrumental in guiding real-time contact tracing efforts during epidemics \cite{firth2020using,kojaku2020effectiveness}. One key limitation of this model is that they are often not spatially or temporally explicit. While this is not problematic when they are used to guide local efforts and forecasts over a short period, this limitation precludes their integration with spatial and temporal forecasts from metapopulation models.}

\blue{Altogether, an integrated modeling framework requires accurate estimation of mobility across regions to inform metapopulation model, as well as individual level data on heterogeneity in contact patterns to guide local interventions. Importantly, these two scales need not be connected in any identifiable way. It is not critical to know the local behavior of individuals connecting different populations. Multi-scale data on the same population, with anonymity assured across scales, would be of itself invaluable improvement to the current reliance on disparate (and often private) data sources on different populations. Doing so could help guide more flexible control policies, beyond blanket measures that ignore the often unknown heterogeneity and adaptive nature of human behavior.}

\blue{More information is therefore necessary to accurately estimate the probability of epidemics given importation event, forecast their eventual spread, and guide control policies. Any particular importation event has the potential to become the initial point of a widespread contagion in a region not reached by epidemics until that point. Since this event basically depends on one individual, aggregated level statistics provide poor description of the heterogeneity of individual level behavior and mobility. Given this uncertainty and the lack of multi-scale real-time informative system, most governments around the world introduced stay-at-home policies to reduce the mobility of their citizens, with the now famous goal of “flattening of the curve.” This expression refers to the process of slowing down the virus spread to keep the need for hospital care below the health care capacity. Different variations of the stay-at-home policy have been introduced, e.g. allowing citizens to move in a predefined radius to walk with their pets and exercise, restricting the number of times per day or the number of people per household that can go out. As previously argued, this is, understandably, a necessary abrupt solution, but one that could be improved by properly utilizing crowd-sourced data and active citizens’ participation. Multi-scale data gathering could inform more flexible restriction while accounting for local heterogeneity in behavior and mobility patterns. We argue that this important data collection efforts could be accomplished by trustworthy federated systems.}

\subsection{Data collection: current activities and potential existing sources}

At higher levels, mining existing data sets collected by different types of service providers is an obvious and powerful way to quantify to which extent citizens follow  \textit{stay-at-home} policies, thereby providing an essential means of assessing their effectiveness. 
For example, smart grid systems can provide information about the changes in the energy consumption in different areas suggesting that people, on average, spend more time at home. The network usage information from broadband operators can also be used as an indicator of change in human behavior related to their mobility. Another source of information are mobility data sets from cellular network operators that are already present on the network. Smart appliances, like smart TVs, smart fridges, and smart light bulbs can be used to provide information about the overall time spent at home versus the time spent outside. Finally, information collected by several smart city applications, such as public transport usage and traffic, could also be leveraged to estimate the change in mobility patterns at the level of places, buildings and vehicles.

At a lower granularity, proximity-tracing platforms have been proposed to collect data at the individual level, for instance the Pan-European Privacy-Preserving Proximity Tracing (PEPP-PT) \cite{pepppt}, often in partnership with mobile service providers and operators. By aggregating proximity data over time, these approaches can follow data protection regulations. Among others, two of the largest Big Tech companies Apple and Google proposed a joint solution that appears to focus  on the protection of user privacy, for example by keeping people anonymous in the central servers and making data submission voluntary.
Different technologies are being proposed to automate and extend contact tracing (e.g. shared databases, GPS traces of confirmed cases, contact tracing through Bluetooth). The use of Bluetooth-based solutions, either in the form of the third-party dedicated apps or as a feature built into the mobile devices' operating systems, seems to be the most promising solution to date.
Although these solutions are a new source of granular data, they cannot cover all relevant forms of infection, namely asymptomatic infections, and infections occurring out of proximity through shared surfaces (e.g. doorknobs).  
In this sense, our hope is to combine the metapopulation and individual tracing models in a citizen science framework to consider more possible routes of infection as well as both proximity (e.g. close contact) and mobility data (e.g. location visited).
In addition to this, machine learning methods are expected to combine expert knowledge (e.g. from virologists and sociologists) with global data collected from different sources for fitting powerful predictive models to high-dimensional data.
In doing so, the target is to leverage the different straights of different data sources at different granularity for privacy protection and allow for more thorough probabilistic forecasts. Ultimately this will allow for graceful lockdowns, tailored incentives for citizens' participation, and fine-tuned legitimate interventions.

\subsection{Data privacy and trustworthiness}

Fine-grained data collection approaches raise two major concerns as highlighted in \cite{cho2020contact}: (1) user privacy; and (2) trustworthiness of the shared information.
Clearly, sharing location and health information with third-party entities can lead to the misuse of the data set for instance through unwanted advertising and health insurance implications.
The impact of fake health information on the contact tracing system should be analyzed to assess its implications in terms of access to testing.  All these applications rely on the premise that users will trust and adopt these applications.
This brings us to an additional problem that could hamper these approaches. The adoption of new technologies is hard to predict, and it might introduce bias for example related to age, wealth, level of education, and state of country development.  
Citizens' participation in warning and response of epidemic outbreaks requires new incentives that reward the responsible use of citizens' personal data for protecting public health, while penalize and prevent citizens' profiling actions, manipulative nudging and power misuse.

\subsection{Challenges and the way forward}

Designing new data fusion schemes tailored to fuel data analytics processes for prevention and mitigation of pandemics is an open research question: Which smart phone sensor data can model compliance of \textit{stay-at-home} policies? How such models can be enhanced with social media, smart grid, or transport data?
Pandemics require that we revisit and potentially reinvent how data should be managed and how systems should be designed to manage data in a more responsible way~\cite{Hagendorff2020}. More specifically, discovering new ways to turn private sensitive data of citizens into public good, while preventing massive surveillance, profiling, and discriminatory analytics, becomes imperative. To pioneer this capability, new transparent and accountable data commons~\cite{Wright2019}, for instance personal data stores~\cite{Alessi2018,Chowdhury2018,Truong2019}, to which citizens have incentives to contribute data, are required. The citizens retain the right for their data to be "forgotten"~\cite{Burkell2017}. Ultimately, they have control and give their consent to how these data are used. These data commons are designed to maximize the use of techniques for privacy preservation~\cite{Vergara2016,Asikis2018}, for example homomorphic encryption, differential privacy, anonymity, and obfuscation, while citizens are made aware of the privacy risks they experience.

\section{Wireless Connectivity and Epidemic-Relevant Information}
\label{sec:wireless}
The previous section covered different aspects related to the data required to support a fine-grained epidemiological model.
It is argued that a ubiquitous system can be employed to manage such data in a privacy-preserving manner.
Here, we identify wireless connectivity and \ac{IoT} devices as a way to collect data following those principles. 

Wireless connectivity and IoT devices come in many flavors, but can, in general, be classified into (1) personal devices, such as mobile phones and earbuds; and (2) unattended connected machines and things, such as surveillance cameras and motion sensors. 
There are three principal sources of epidemic-relevant data acquired through wireless connectivity: (1) online social networks; (2) personal smart phone and mobile data; and (3) sensory and Internet of Things (IoT) devices.

\subsection{Online social networks}

In critical events, people tend to use online social networks to post comments about the emergency and learn from other users' comments. As a result, online social networks become a rich source of diverse information that could help to understand the main characteristics of the crises and their potential magnitude in early stages.
For instance, in  \cite{9043580}, authors collected comments posted in the Chinese social media channel Weibo during the first month of the COVID-19 epidemic in Wuhan (China) to understand the evolution trend of the public opinion. A similar approach was used on Twitter in  \cite{alshaabi2020world,socialnetwork_covid_Rosa2020}. Government entities can use this approach to give more attention to the needs of the public during the beginning of the epidemic and adjust their responses accordingly.
Another form of data content, normally provided by explicit consent from the citizens, is the GPS location data from personal devices. At the application level, the Twitter platform offers various user information, geographical location being one of the most important for a wide range of applications. However, only a few users enable their geolocation as public. To discover the user's geolocation (geographical region or exact geocoordinates), there are mainly two approaches called content-based and network-based algorithms. The first one uses textual contents from tweets, and the latter uses the connections and interactions between users. In \cite{8462191}, a neural network model based on multiview learning by combining knowledge from both user-generated content and network interaction is proposed to infer users’ locations.

\subsection{Personal smart phone devices}

This source of information comes in many different forms, ranging from the use of metadata in cellular mobile networks for user localization  up to the metadata associated with different applications, such as Twitter. 
The use of data\footnote{They are referred to as metadata due to its role played in the communication system to which it belongs; here we will also refer to it as a ``data'' when there is no danger of ambiguity.} 
collected in mobile networks during the COVID-19 pandemic has 
received significant attention \cite{Oliver2020MobileNet}. Agglomerated mobile network 
data can be used to verify if interventions, such as school closures, are effective, and help to understand the geographical spread 
of an epidemic. Different types of mobile network metadata are collected at different
levels of the communication system, and they offer varying levels of information. At the lowest level, localization of
mobile network users is possible by 
evaluating the strength of the wireless signal received at base stations. Depending on how many base stations are connected to the device, the location can be determined at the level of entire cells or down to a few meters using triangulation. 

The resolution offered by cellular mobile localization maybe too coarse for detecting citizens' proximity and potential infections by face-to-face interactions. 
Nevertheless, even low-resolution location data can give insight into behavioral 
patterns (how much time is spent at home, office, and events) of individuals. Proximity detection can be enhanced by using the metadata from Bluetooth devices, such as beacons and discovery messages (e.g., \cite{ferretti2020quantifying}).
Wi-Fi is also a technology that is used ubiquitously, and its metadata can provide information on user proximity. For example, proximity can be inferred by comparing the lists of access points that each device can see within a given time interval \cite{sapiezynski2017inferring}. 

Fusing the metadata from these different sources, along with the context information (how many family members are at 
home, which events) is very relevant for monitoring the epidemic. 

\subsection{Sensory and IoT devices}

Many applications and devices become an important tool to provide epidemic-relevant data.
These can be related to surveillance/thermal cameras, drones and even wearable devices.
For example, surveillance cameras can be employed to count the number of people entering and exiting a specific area.
In other scenarios, thermal cameras are used with specialized settings to focus on human skin temperature range, and infrared spot sensors for individual temperature scanning.
A similar application to identify individual or group activities in a given place is based on motion sensors as infrared lasers, or ultrasonic sensors.
In this approach, the motion sensors are installed and deployed in a specific environment to recognize different human activities.

Other important source of \ac{IoT} data are the smart healthcare devices, for instance, remote monitoring systems to check body temperature, which is a key sign in the support of homeostasis.
Other popular applications are oxygen saturation monitoring based on beat oximetry, electrocardiogram monitoring with a specialized framework to estimate the heart rate, elderly monitoring using doppler radar  to identify risk movements of elderly people, sugar level monitoring, and blood pressure monitoring.
These \ac{IoT} devices provide information that is analyzed by the \ac{IoT} data to support specific applications.

\section{Federated Global Epidemiological Warning System }

\begin{figure*}[!t]
\centering
\includegraphics[width=0.9\textwidth]{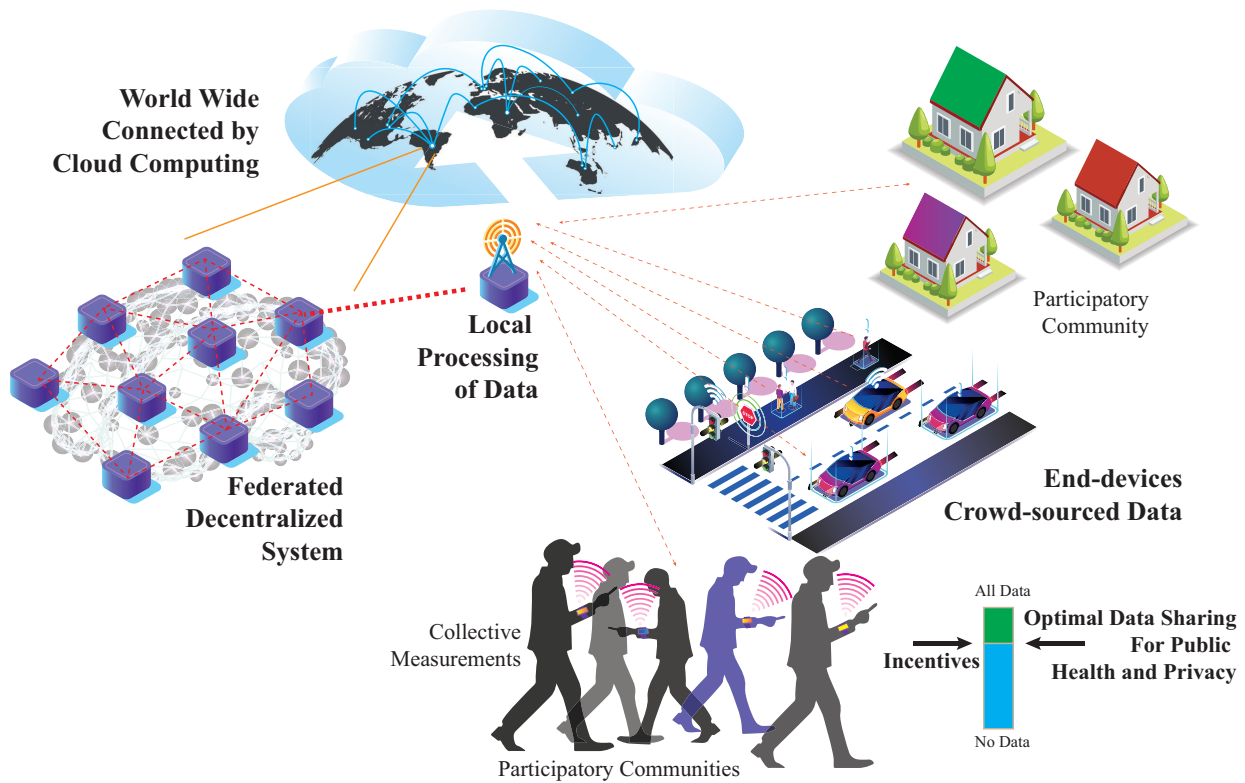}
\caption{A federated ubiquitous systems for epidemiological warning and response. An organic and bottom-up scaling at global level is envisioned based on active citizens' participation. Decentralized privacy-preserving computations are performed from the the edge to the cloud based on crowd-sourced obfuscated and anonymized data managed with distributed ledgers to empower trust. Incentive mechanisms for responsible data sharing align the public health mandate with citizens' privacy and autonomy.}
\label{fig:system_model}
\end{figure*}

We envision a federated and decentralized coordination system for epidemiological warning and response. This system is federated by citizen communities that crowd-source data, personal smart phone devices, community-level IoT devices (e.g. LoRaWAN networks) and other computational resources on the edge. It can scale organically and bottom-up to city-level, national-level and ultimately at global-level to coordinate in a socially responsible way the international actions of public health organizations and governments. This scaling requires tailored incentives that align public health policy-making with citizens' privacy and autonomy. Distributed ledgers with secure crypto-economic incentive models running on edge serves and scaling-up on-demand at a global level using wireless connectivity and public cloud infrastructure are the means to support the federated nature of this proposed system. Figure \ref{fig:system_model} illustrates the our vision.

\label{sec:proposal}
\subsection{Key components} 
The proposed federated global epidemiological warning and response system is built upon epidemic-relevant information obtained through wireless connectivity introduced in Section \ref{sec:wireless}, and the internet infrastructure including fiber-optic and satellite links.
Thus, in addition to the devices that are the data sources (e.g., smart phones, wearables, smart appliances, and cameras) with their applications and existing communication networks, there is a need for a computer infrastructure dedicated to store and process the epidemic-relevant data in a secure and privacy-preserving manner.
It is argued that this infrastructure also needs to follow a similar federated organization based on the geographic locations.
Each municipality, city, county, or larger neighborhood will rely on edge servers to process their respective data contents.
The premise is to keep the computations as local as possible.

However, larger-scale computations related to the interrelations between locations (e.g. mobility, commuting, and traveling) requires data from these different places. 
Once again, the federated organization supports this collaborative sharing (up and downstream between the federated entities), but it will probably require computationally more demanding algorithms. 
It is possible that those computational tasks require more powerful cloud servers, or collaborative parallel computing at the edges.
For security reasons, \acp{DLT} shall be employed to store data from the edge servers at the global level.
At the technical level, the following aspects deserve special attention.

\subsubsection{Privacy}

In this paper, a privacy-by-design approach is proposed with which several critical operations of the federated global epidemiological system can be performed; for instance, decentralized data analytics~\cite{Bonawitz2017,Pournaras2017}, social interactions analysis~\cite{Musciotto2016,Bahri2018}, decentralized planning and resource allocation~\cite{Pournaras2018} and federated learning~\cite{Yang2019,Smith2017}. Such operations integrate in a smart way state-of-the-art techniques, for instance, informational self-determination, homomorphic encryption, differential privacy, obfuscation, and anonymity. 
Often, privacy may limit the quality of service known as the \emph{system utility} as the accuracy and quality of data are deteriorated to hide information content. For instance, predicting the risk of infections for individuals based on 
an epidemiological network model is a graph-based semi-supervised learning problem. Privacy-preserving semisupervised learning over graphs has been considered in \cite{Arai2011}. However, the precise trade-off between (lowering of) privacy protection and learning accuracy is largely open.
Pareto optimal trade-offs can be configured and regulated by tuning the parameters of the privacy techniques as previously shown in~\cite{Asikis2018}. Monetary and other incentives can be used to coordinate data sharing choices in a crowd. However, new types of social incentives are required for such an epidemiological system; for instance, incentives related to well-being, receiving solidarity, and long-term payoffs.

\subsubsection{Interoperability}
The big data required by the proposed solution have to be interoperable, i.e. the several applications that are providing data to be used to accomplish a specific task have to operate together and share a common "understanding" \cite{Jacoby2017}. This can be achieved by employing standards for health informatics such as ISO TS22220:2011, TS21090:2011, or TS13606. For instance, the data collected from different sources can be used to predict refinements to patient care or new drug contraindication \cite{Schulz2019}.
This key issue, though, has deserved little attention in large-scale epidemiological studies \cite{roos2017preparing, moreira2016improving};
it is usually assumed that heterogeneous data sources are compatible with each other.
In practice, though, the highly heterogeneous data sources lead to poor interoperability, which creates barriers to effectively combat pandemics like COVID-19, as indicated by \cite{legido2020high}.

\subsubsection{User interfaces}
A successful platform also involves a suitable end-user interface \cite{maciejewski2011pandemic}. 
In this sense, data consumption by public health officials and global health agencies will require user-friendly web-based interfaces, using common dashboarding techniques.
However, beyond this, the proposed federated platform has to consider citizen participation and the heterogeneity of end-users.
Therefore, the following characteristics should be taken into consideration in the design of the platform: it should be (a) \textbf{explainable/accountable} to improve for instance awareness and engagement ~\cite{Genevieve2019}; (b) \textbf{gamified} to engage and incentivize participation~\cite{Neto2020,Wazny2018}; and (c) \textbf{customizable} for different user groups at an international level.

\subsection{Proposed architecture}

The proposed architecture has to articulate the key elements from data acquisition to analytics, following the best practices related to privacy and cyber-security.
At the acquisition level, in addition to the existing data retrieval from the web, wireless-connected devices will send data to edge nodes that will be associated with specific regions. 
Data will be anonymously preprocessed at the edge, including some intelligent detection of anomalies or event detection, and then sent to regional (cloud) servers to run a more complete model that will fuse the geolocated timestamped data and run detection models based on explainable AI approaches combined with mathematical computational methods.
In this case, crowd-sourcing models are applied based on hardware as a public good in an approach similar to the DIASPORA social network \cite{diaspora}, also extending this approach to software and data. 
Regional models will be associated in \textit{federations} resembling the governance structure of the actual regions under interest, which is customizable to local policies/governance models.

Following this federated and collaborative organization, it will be possible to build organically a \textit{fine-tuned early detection} system at the global level in a decentralized yet hierarchical manner to support \textit{graceful lockdowns}. 
This would turn the proposed federated architecture into a holarchy, where each level is independent and self-sustained, but it can also be encapsulated at a level above to capture new goals~\cite{diaconescu2018hierarchical}.

This system will rely on \acp{DLT} to guarantee a trustful system without a responsible third party and minimize the risk of data manipulation\cite{dltnbiot}. 
Distributed ledgers could also be used to facilitate different (crypto-economic) incentive mechanisms, such as token curated registries.
Its main objective is to build a crowd-sensing trustworthy platform based on privacy-preserving methods to detect potential harmful symptoms in a specific region in almost real-time and flag them to relevant authorities.
This alarm needs to be accurate, reliable, and explainable.
It will also require a user-tailored interface that could empower citizens by providing detailed explanations and easy-to-access information, while being a tool for policymakers (from the city level to global organizations) designing the correct interventions (a variety of incentives, sanctions, and other persuasive measures) given the specific context of the epidemic in some specific region in a given period of time.

\subsection{Representative Use Case}
To obtain insights from a practical implementation, we define a specific use case that matches the premises of the proposed architecture. This key industry use case focuses on\textbf{ remote patient monitoring}. Here, many issues in treating chronic patients could be reduced or resolved through more efficient patient care. In the case of pandemics, such as COVID-19, the remote monitoring of \blue{diagnosed} patient can represent a key difference to avoid the virus acceleration. Besides that, the application should guarantee privacy and security on the gathered data. To guarantee people participation, the system can provide incentives, such as bonus or tax compensations. 
In the context of the proposed federated global epidemiological warning system, the remote patient monitoring is a representative use case, in which the integration between DLTs and IoT devices plays a key role. We analyze this integration through an experimental setup based on NB-IoT, which is one of the most representative \ac{LPWAN} technologies in cellular networks. Considering that the proposed architecture requires massive end devices deployment, low power consumption, superior coverage, and worldwide deployment, NB-IoT becomes a key solution to support the federated global epidemiological warning system.
Some details of the integration setup are described in the next sub-section, emphasizing data workflow, and end-to-end (E2E) latency. We consider latency because the experimental results indicate that it is the most sensible network parameter when DLTs are built in the top of the connectivity system.

\subsection{Experimental Setup and Results}
\label{sec:experim-results}
\subsubsection{Setup Description}
\begin{figure}[!t]
\centering
\includegraphics[width=0.5\textwidth]{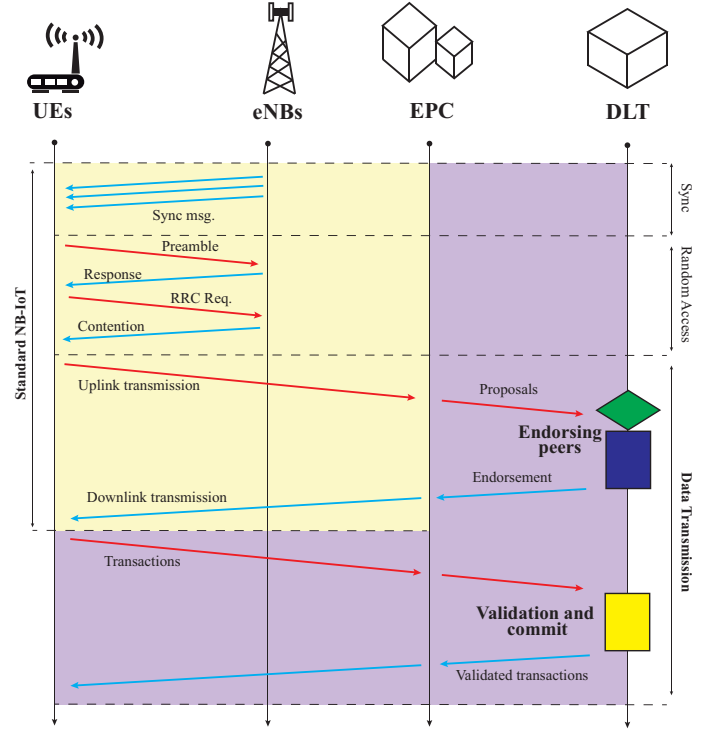}
\caption{Data workflow in both scenarios: standard \ac{NB-IoT} in yellow background, which is complemented by DLT-based system in purple background.}
\label{fig:ldt_workflow}
\end{figure}

In this section, we provide brief explanation of the integration between the \ac{NB-IoT} network and the DLT-based system, which is illustrated in Figure \ref{fig:ldt_workflow}.

In a conventional \ac{NB-IoT} system, the uplink data generated by the \acp{UE} is transmitted though specialized packet messages and routed toward an edge data center to be stored and processed (yellow background in Fig. \ref{fig:ldt_workflow} - representing a standard-NB-IoT). 
At this point, the monitoring system has no control on the collected data, so modification, corruption, and losses may occur. 
Conversely, in our DLT-enabled \ac{NB-IoT} setup, the uplink data generated by the \acp{UE} is transmitted to a randomly chosen group of endorsing peers of Hyperledger Fabric, a type of DLT used in our study, as transaction proposals (check the purple background in Fig. \ref{fig:ldt_workflow}). 
Then, each of the peers signs the transaction using the elliptic curve digital signature algorithm (ECDSA) and adds the signature before returning the signed message back to the IoT devices (\acp{UE}).

The peers that provide an endorsement of the ledger send an update to the application, but do not immediately apply the proposed update to their copy of the ledger. 
Instead, a response is sent back to the IoT device to confirm that the transaction proposal is correct, has not been previously submitted to the ledger, and has a valid signature. 
Therefore, the security increases with the number of endorsing peers. In addition, smart contracts can be executed to update or query the ledger. 

Then the IoT device broadcast the confirmed transaction proposals along with the confirmation to the ordering service. The received transactions are ordered chronologically to create blocks. 
These transaction blocks are delivered to all peers for validation. Then, the peers append the block to the ledger, and the valid transactions are committed to the current state database. 
Finally, a confirmation message is transmitted back to the IoT devices to report that the submitted transaction has been immutably published to the ledger. The \blue{importance} of confirmation is analyzed in \cite{feedback}.

\subsubsection{Results}
We evaluate the performance of an integrated \ac{DLT} and IoT system with \textit{permissioned Hyperledger} Fabric and \ac{NB-IoT}. 
Our experimental setup is based on Hyperledger Fabric v1.4, \ac{NB-IoT} development kits SARA EVK N211 and one NB-IoT Amarisoft eNB station. 
We compare the E2E latency of conventional \ac{NB-IoT} with DLT-based \ac{NB-IoT} in various scenarios. The number of endorsing peers is varied from 1 to 4 peers. 
The IoT device transmits packet every minute, and the total number of messages transmitted is around 1000.

\begin{figure}[!t]
\centering
\includegraphics[width=0.55\textwidth]{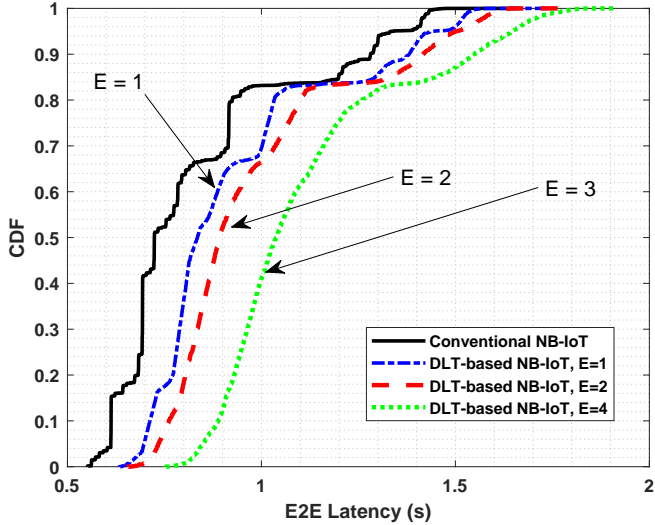}
\caption{End-to-end latency performance of the proposed integration between \ac{DLT} and \ac{NB-IoT} with different number of endorsed peers.}
\label{fig:ldt_result}
\end{figure}

Figure \ref{fig:ldt_result} shows that the end-to-end (E2E) latency is computed based on the total uplink and downlink transmission latency and validation latency in the ledger.
We observe that when adding an new endorsing node to the same channel, this will increase the endorsing latency.
In comparison with conventional \ac{NB-IoT} system, the DLT-based \ac{NB-IoT} perform a higher latency due to the validation process from ledger to verify transaction as well as building block. 
This is a trade-off between DLT-based system and conventional IoT system. DLT-based IoT system guarantees the trust; however, the latency is slightly degraded.

\section{Discussion and Future Perspective}
\label{sec:concl}
\subsection{Latency analysis on practical DLT and NB-IoT Integration Setup}

A federated global epidemiological warning system was proposed. Based on it, a specialized use case focused on remote patient monitoring was considered. As one critical piece of this use case is the integration of DLTs and IoT devices, we implemented an experimental integration setup between \acp{DLT} and \ac{NB-IoT}.

In this study, we conclude that the latency increased when the number of endorsing peers is increased.  Results indicate that in average the latency is increased  0.25 seconds per endorsed peer that is added.
However, this increases the guarantees in terms of secrecy, privacy, and interoperability. 

\blue{The experimental setup also reveals additional challenges}, such as the performance of DLT-based system with massive number of connections, \blue{as well as a trade-off between privacy, interoperability and system performance.}

\subsection{Federated Global Epidemiological Warning System}
Although the proposed architecture is designed to be fair among all citizens, few important practical points still deserve attention.
Good quality data usually follow the wealth distribution over all scales of the proposed federation. 
In other words, good quality data is very likely to be less available in poorer regions of a city.
The same is valid for poor regions within a country and poor countries.
This existing gap needs to be considered when building incentives for participation, otherwise the proposed solution has the potential of reinforcing inequality \cite{gianfrancesco2018potential}.

Another current issue we cannot neglect is the public perception of wireless technologies.
In this sense, part of the general public perceives IoT as insecure \cite{zubiaga2018longitudinal} and \ac{5G} as a health hazard \cite{hasan20195g} to the point of claiming an astonishing causal relation between 5G and COVID-19, which has caused destruction of base stations across UK.
Beyond these conspiracy theories that are hard to combat, there exist legitimate concerns of anonymized data not being anonymous \cite{rubinstein2016anonymization}, and of novel surveillance techniques introduced in times of crisis that are maintained for monitoring (legally or illegally) populations, e.g. post-9/11 surveillance in the USA \cite{landau2013making}.

All these have to be carefully addressed from the early stages of the system development. Furthermore, it has to be clear that neither \emph{technology} nor \emph{data} can prevent another outbreak on their own, but can only provide the extremely valuable tools to enable the holy grail of controlling epidemics: \textit{early detection, early response} along all relevant actors within the federated organization.
In other words, technology and data detect and identify potential harms and suggest actions and reactions, but the final diagnostic and further interventions are due to the responsible institutions within the federated structure and citizens' active participation.
Our proposal answers those challenges based on informational self-determination as the way to build trustworthy and secure public infrastructure that shall enable \textit{graceful lockdowns} as advocated here.
\textbf{In this sense, the proposed solution introduces a more balanced management strategy, moving away from purely top-down approaches toward a participatory system where the citizens are active.
We further expect that, even without being implemented, the high-level architecture introduced here could offer important technological suggestions to decision-makers of how to start smoothly resuming activities after lockdowns.}

\section{Conclusions}
The federated system warning response envisioned for epidemiological warning and response is a potential enabler to scale a global-level coordination between public health organizations and governments. 
As it is based on \acp{DLT}, secure crypto-economic incentive models can be used to motivate citizens and institutions.
Simultaneously, privacy and interoperability is assured.  

An specific use case of the proposed system is based on the integration of \acp{DLT} and a popular \ac{IoT} transport technology, such as \ac{NB-IoT}.
A laboratory setup implementation of this integration was proposed and detailed in this paper. 
It indicated  that there is a trade-off between privacy and some key network \acp{KPI}, such as latency. This trade-off should be studied and analyzed in terms of interoperability and system performance in order to maximize the benefit of the proposed trustworthy federated system. 

\bibliographystyle{IEEEtran}
\bibliography{main.bib}

\end{document}